\documentclass[12pt]{iopart}
\usepackage[pdftex]{graphicx,color}
\usepackage{amssymb}
\usepackage{epsfig}
\usepackage{subfigure}
\usepackage{textcomp}
\usepackage{cite}
\usepackage{float}
\usepackage{bm}
\expandafter\let\csname equation*\endcsname\relax
\expandafter\let\csname endequation*\endcsname\relax
\usepackage{amsmath}
\graphicspath{ {./figs/} }

\begin{document}
\title{Role of Depletion on the Dynamics of a Diffusing Forager}

\author{O. B\'enichou} \address{Laboratoire de Physique Th\'eorique de la
  Mati\`ere Condens\'ee (UMR CNRS 7600), Universit\'e Pierre et Marie Curie,
  4 Place Jussieu, 75252 Paris Cedex France}

 \author{M. Chupeau} \address{Laboratoire de Physique Th\'eorique de la
  Mati\`ere Condens\'ee (UMR CNRS 7600), Universit\'e Pierre et Marie Curie,
  4 Place Jussieu, 75252 Paris Cedex France}
  
\author{S. Redner} \address{Santa Fe Institute, 1399 Hyde Park Road, Santa
  Fe, New Mexico 87501, USA}

\begin{abstract}

  We study the dynamics of a \emph{starving random walk} in general spatial
  dimension $d$.  This model represents an idealized description for the fate
  of an unaware forager whose motion is not affected by the presence or
  absence of resources.  The forager depletes its environment by consuming
  resources and dies if it wanders too long without finding food.  In
  the exactly-solvable case of one dimension, we explicitly derive the
  average lifetime of the walk and the distribution for the number of
  distinct sites visited by the walk at the instant of starvation.  We also
  give a heuristic derivation for the averages of these two quantities.  We
  tackle the complex but ecologically-relevant case of two dimensions by an
  approximation in which the depleted zone is assumed to always be circular
  and which grows incrementally each time the walk reaches the edge of this
  zone.  Within this framework, we derive a lower bound for the scaling of
  the average lifetime and number of distinct sites visited at starvation.
  We also determine the asymptotic distribution of the number of distinct
  sites visited at starvation.  Finally, we solve the case of high spatial
  dimensions within a mean-field approach.

\end{abstract}

\maketitle

\section{Introduction}

The \emph{starving random walk} represents an idealized description for the
dynamics of a forager without any sensory awareness who randomly searches for
food in an environment that is depleted by the foraging process
itself~\cite{BR14}.  In this model, the forager undergoes an unbiased
nearest-neighbor random walk on a regular $d$-dimensional lattice, in which
each lattice site initially contains one food unit.  The walk has a
metabolic capacity, or intrinsic starvation time $\mathcal{S}$, which is
defined as the number of steps that the walk can travel without
encountering food before starving to death.  Each time the walk lands on an
empty site, it comes one time unit closer to starvation.  However, if the
walk lands on a site that contains food, the food is instantaneously and
completely consumed, and the walk can again travel $\mathcal{S}$ additional
steps without encountering food before starving.

Our goal in this work is to quantitatively account for the phenomenon of
starvation, in which an organism fails to find resources before it expends
its own metabolic reserves and starves to death.  This search for resources
is an essential task of all living
organisms~\cite{C76,B91,OB90,KM01,ASD97,KR85,SK86,Vea96,LKW88}, in which the
resource could be nourishment, an abode, or a particular individual.  Here we
view the resource as food that is gradually depleted through consumption by
foragers; the ecologically relevant situation where the resource is
replenished has also been investigated~\cite{CBR16}.  Search for a resource
has been extensively studied for an omniscient forager that has full
knowledge of its environment (see, e.g., \cite{C76,B91,OB90}).  In this
setting, the basic objective is to determine the criterion that optimizes the
foraging process.  Our focus is on the complementary case where the forager
has no environmental knowledge nor the ability to learn about its environment.
Because the resource is gradually depleted by the forager, the ultimate fate
of the forager is to necessarily starve.

An important aspect of the starving random walk is that it moves in the same
manner, irrespective of its distance to food sources, as well as whether or
not it has recently encountered food.  This feature contrasts with the
\emph{excited} random walk~\cite{PW97,D99,P07,BW03,ABV03,Z04,AR05}, in which
the walk continues to move in the same direction when it encounters food
and wanders randomly when it does not.  While the excited random walk
exhibits surprising properties, the starving random walk is also quite
phenomenologically rich.  The salient feature of the starving random walk is
that its mortality is coupled to its gradually depleting environment.  In
this sense the starving random walk also differs from \emph{mortal} random
walk models~\cite{LK09,BNHW87,YAL13,AYL13}, in which a walk spontaneously
dies at a fixed rate, independent of its trajectory.

We will investigate two basic questions: (i) How long does a forager live?
(ii) What is the spatial extent of the region explored by a forager before it
starves?  The behaviors of two quantities depend in an essential way on the
metabolic capacity $\mathcal{S}$ and the spatial dimension $d$, as previously
discussed in~\cite{BR14}.  Here, we provide details of the derivation of
these results, as well as qualitative arguments to support our analytical and
numerical results.

We will first treat the case of one dimension, where the problem is amenable
to a full analysis (Sec.~\ref{sec:1d}).  For any $d\leq 2$, the forager tends
to carve out a compact depleted spatial region---which we define as the
``desert''---because of the recurrence of the random walk~\cite{F68,W94}, and
then starve before it can escape this desert.  As a consequence, the forager
lifetime grows slowly with $\mathcal{S}$---linearly in $\mathcal{S}$ for
$d=1$ and as $\mathcal{S}^z$ with $z\approx 1.9$ for for the ecologically
relevant case of $d=2$, as will be presented in Sec.~\ref{sec:2d}.  We also
introduce a circular approximation to provide a lower bound for the lifetime
and number of sites visited at the instant of starvation in $d=2$.

As the spatial dimension increases, the transience of the random walk implies
the forager is less likely to return to previously depleted sites.
Consequently, the lifetime of the forager should be an increasing function of
the dimension for a fixed metabolic capacity ${\cal S}$.  Thus a starving
random walk is long-lived for $d>2$; simulations suggest that the average
survival time grows as $\exp(\mathcal{S}^\omega)$, with
$\omega\approx \frac{1}{2}$ in $d=3$ and with $\omega$ a gradually increasing
function of the spatial dimension~\cite{BR14}.

In Sec.~\ref{sec:infd}, we present a mean-field approximation in which
successive visits to food-containing sites are uncorrelated; this description
corresponds to the starving random walk model in the limit of $d=\infty$.
Here, we solve for the first-passage probability for the walk to starve,
from which we find that the average forager lifetime grows exponentially with
$\mathcal{S}$.  We close with some concluding remarks in
section~\ref{sec:conc}.  Calculational details for the distribution of
visited sites and the average lifetime in one and two dimensions are
presented in the Appendices.

\section{One Dimension}
\label{sec:1d}

\begin{figure}[ht]
\centering
\includegraphics[width=0.55\textwidth]{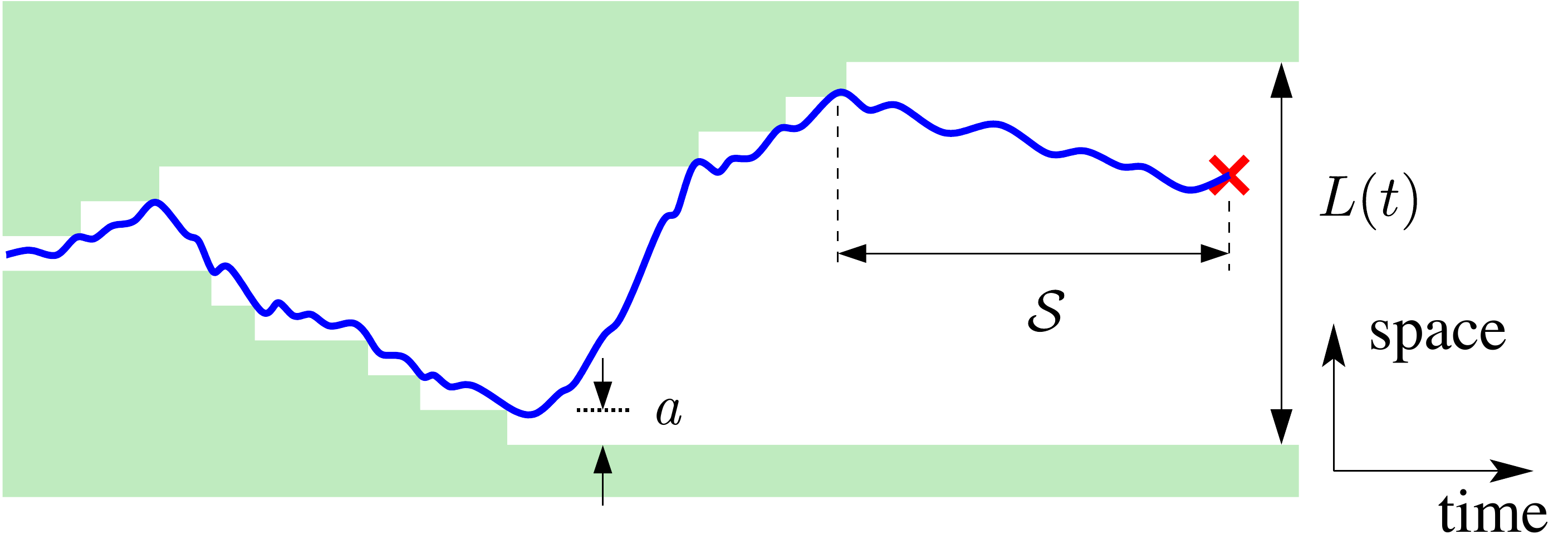}
\caption{Space-time evolution of a one-dimensional random walk that clears
  out an interval---the desert---where food (shaded) has been eaten.
  Whenever the walk reaches a food-containing site, the food is consumed
  and the desert grows by one lattice spacing $a$.  The walk starves
  (\textcolor{red}{$\boldsymbol{\times}$}) when it travels $\mathcal{S}$
  steps without encountering food.}
\label{interval}  
\end{figure}

The evolution of the system in one dimension is schematically illustrated in
Fig.~\ref{interval}.  As the walk moves, a region of length $L(t)$ that is
devoid of food---the desert---is carved out.  The survival of the walk is
controlled by the interplay between its wandering and going hungry in the
desert interior and reaching the edge of this desert to consume food.  We are
interested in the average of the lifetime of the walk and of the number of
distinct sites visited by the walk, as well as the distribution of the latter at the
moment of starvation.  Because of the simplicity of one dimension, these
properties can all be calculated explicitly.  It is instructive, however, to
begin with an intuitive argument for the lifetime and number of distinct
sites visited.

\subsection{Heuristic approach}
\label{heuristic}

In one dimension, the walk clears out a desert whose length grows
sporadically by steps of a single lattice spacing $a$ (Fig.~\ref{interval}).
We estimate the average walk lifetime by partitioning a typical trajectory
into three stages:
\begin{enumerate}
\item The walk first carves out a desert of a ``dangerous'' critical
  length $L_c$---duration $T_1$.
\item The walk returns $M$ times to the same edge of the desert---duration
  $T_2$.
\item The walk wanders for too long inside the desert and dies---duration $T_3$.
\end{enumerate}

During stage (i), the walk successively reaches either edge of the desert
and extends it by one lattice spacing as long as the average crossing
time
\begin{equation}
\label{cross}
t_{\rm cross} = \frac{L^2-a^2}{6D}
\end{equation}
is less than the metabolic capacity $\mathcal{S}$.  Equation~\eqref{cross}
applies for a walk that starts a distance $a$ from one edge of a desert of
length $L$ and reaches the other side~\cite{R01}.  We define the critical
desert length $L_c \equiv n_c \, a$ by the condition that the walk
typically starves when it attempts to cross the desert.  This gives
$t_{\rm cross}=\mathcal{S}$, or
\begin{equation}
L_c \simeq \sqrt{6D\mathcal{S}}\,
\end{equation}
in the large-$\mathcal{S}$ limit.  When the length of the desert equals $ka$,
the average time $t_k$ needed to reach \textit{either} edge of the desert when
starting a distance $a$ from one edge is~\cite{R01}
\begin{equation}
t_k=\frac{(k-1)a^2}{2D}\,.
\end{equation} 
Thus the duration $T_1$ of the first stage of a typical trajectory is the sum
of these average return times $t_k$ until the desert reaches the critical length
$L_c\equiv n_c \, a$:
\begin{equation}
T_1 \simeq \sum\limits_{k=1}^{n_c} \frac{a^2 k}{2D} \simeq \frac{L_c^2}{4D} 
= \tfrac{3}{2} \mathcal{S}\,.
\end{equation}

During stage (ii), the walk likely starves if it attempts to cross the
desert.  Thus it continues to survive only if it makes repeated excursions to
the same side of the desert.  Suppose that the walk is at the left edge at
the end of stage (i).  The walk will typically return $M$ times to this
edge before attempting to cross the desert, which results in its starvation.
The probabilities that the walk \emph{eventually} hits the left edge or the
right edge of a desert of length $L$ when starting a distance $a$ from the
left edge are, respectively,~\cite{R01}
\begin{equation}
p_- =\frac{L-a}{L}\qquad\qquad\qquad\qquad\qquad p_+ = \frac{a}{L}\,\,,
\end{equation}
and the conditional first-passage time to the left edge is
\begin{equation}
\label{t}
t_-= \frac{aL}{6D}\Big(2-\frac{a}{L}\Big)\,.
\end{equation} 
The probability $\Pi_k$ that a random walk hits the left edge $k$ times
before hitting the right edge, starting from a desert of length $L_c$,
therefore is
\begin{align}
  \Pi_k
  &=\frac{L_c-a}{L_c}\times\frac{L_c}{L_c+a}\times\frac{L_c+a}{L_c+2a}\times\ldots\times\frac{L_c+(k-2)a}{L_c+(k-1)a}\,\,\times\,\,\frac{a}{L_c+\!ka} \nonumber \\
  &=\frac{a(L_c-a)}{[L_c+(k-1)a](L_c+ka)}\,.
\end{align}
The first $k-1$ terms in the product give the probabilities for the walk to
successively hit the left edge; after each such event the interval grows by
$a$.  The last factor is the probability to hit the right edge when the
interval has reached length $L_c+ka$.  It is straightforward to verify that
this probability distribution is normalized, $\sum_{k\geq 0}\Pi_k=1$.
Because $\Pi_k\sim k^{-2}$ for large $k$, the average number of same-side
excursions until the right edge is reached, $\sum_{k\geq 0}k\, \Pi_k$, is
infinite.  Nevertheless, the \emph{typical} number $M$ of same-side
excursions is meaningful.  This quantity can be defined as the median number
of such excursions, so that $\sum_{0\leq k\leq M}\Pi_k=1/2$.  This gives
\begin{equation}
  \sum_{0\leq k\leq M} \frac{a(L_c-a)}{(L_c+(k-1)a)(L_c+ka)}
=  1-\frac{L_c-a}{L_c+Ma} =\frac{1}{2}~.
\end{equation}
Thus for an interval of length $L_c$, the right edge will typically be
reached after $M\sim L_c/a$ consecutive same-side excursions to the left
edge.  This condition defines the end of the stage (ii).  Its duration $T_2$
is the sum of the average conditional return times to the same edge of the
desert when starting one lattice spacing away from this edge:
\begin{equation}
T_2 \simeq \sum_{k=L_c/a}^{L_c/a+M} \frac{k a^2}{3D} \simeq \frac{L_c^2}{2D} 
= 3 \, \mathcal{S}\,.
\end{equation}
Finally, in stage (iii), the walk fails to return to the left edge of the
desert within $\mathcal{S}$ steps.  The duration $T_3$ of this terminal stage
is exactly $\mathcal{S}$ steps.

We therefore estimate of the average lifetime $\tau$ of the starving random walk
as
\begin{equation}
\tau \simeq T_1+T_2+T_3 \simeq 5.5 \, \mathcal{S}
\end{equation}
compared to the exact asymptotic value of 3.27686\ldots $\mathcal{S}$ that
will be derived in the next section (see also~\cite{BR14}).  Note that $T_1$,
$T_2$ and $T_3$ are comparable when ${\cal S}\to \infty$.  Moreover, the
above heuristic approach gives the average number of distinct sites visited
$\langle \mathcal{N} \rangle$ as
\begin{equation}
\langle \mathcal{N} \rangle \simeq \frac{L_c}{a}+M \simeq 2 \frac{\sqrt{6D\mathcal{S}}}{a} \simeq 3.4641 \, \sqrt{\mathcal{S}}\,,
\end{equation}
compared to the exact asymptotic value of 2.9022\ldots $\sqrt{\mathcal{S}}$~\cite{BR14} that
we will also derive in the next section.

\subsection{Exact probabilistic approach}

We now calculate the average number of distinct sites visited by the walk
and its distribution at the instant when the walk starves, as well as its
average lifetime.  We then provide explicit results by an asymptotic
analysis.

For an unbiased $n$-step nearest-neighbor random walk in one dimension, the
average number of distinct sites visited, which is also the span of the walk
divided by the lattice spacing $a$, asymptotically grows as
$S_n\sim \sqrt{8n/\pi}\approx \sqrt{2.55\,n}$~\cite{W94}.  The new feature
for starving random walks is that there cannot exist excursions of more
than $\mathcal{S}$ consecutive steps in the previous history in which the
walk does not encounter food.  Thus a long-lived starving random walk
spends less time roaming within the interior of a desert than an unrestricted
random walk.  As a consequence, the number of distinct sites visited by
starving random walks should be larger than that for unrestricted random
walks at the same time.

To determine the number of distinct sites visited and its distribution at the
instant of starvation, we define $V(\mathcal{N})$, the probability that a
random walk has visited $\mathcal{N}$ distinct sites when starvation
occurs.  We may express this probability as
\begin{equation}
\label{PLdef}
V(\mathcal{N})= \mathcal{F}_2 \,\mathcal{F}_3\, \mathcal{F}_4 \ldots \mathcal{F}_\mathcal{N} (1-\mathcal{F}_{\mathcal{N}+1})\,,
\end{equation}
where
\begin{equation*}
\label{pn}
\mathcal{F}_k=\int_0^\mathcal{S} dt\, F_k(t)\,,
\end{equation*}
with $F_k(t)$ the first-passage probability to either edge of an interval of
length $ka$ when the walk starts a distance $a$ from one edge.  That is,
an interval of length $ka$ has to be generated, then an interval of length
$(k+1)a$, then $(k+2)a$, etc., each within a time $\mathcal{S}$, until the final
interval length is reached.  

The details of this derivation are given in \ref{P}; we make contact with the
discrete nearest-neighbor random walk by setting the diffusion coefficient
$D=a^2/2$.  From the asymptotic result given in Eq.~\eqref{Q}, the average
number of visited sites at the starvation time is, in the large ${\cal S}$
limit,
\begin{align}
\label{N}
 \langle \mathcal{N}\, \rangle = \sum_{ \mathcal{N}\geq 1} \mathcal{N}
  V(\mathcal{N})
  & \simeq \mathcal{N}^*\int_0^\infty \theta \,V(\theta)\, d\theta\,,
\end{align}
where we take the continuum limit and express the distribution of visited
sites in terms of the scaled variable $\theta = \mathcal{N}/\mathcal{N}^*$,
with $\mathcal{N}^*=\pi\sqrt{D\mathcal{S}}/a$.  Using the explicit form for
the distribution $V(\theta)$ in Eq.~\eqref{Q} and computing this integral
numerically gives
\begin{equation}
\label{Nav}
\langle\mathcal{N}\rangle \simeq 1.3065\,\, \mathcal{N}^*\approx 2.9022\, \sqrt{\mathcal{S}} 
\qquad\qquad {\cal S} \gg 1.
\end{equation}

We now turn to the average lifetime $\tau$, which we write as
\begin{equation}\label{tausum}
\tau = \sum_{n=0}^\infty
\big(\tau_1+\tau_2+\dots +\tau_n+\mathcal{S}\big) V(n)\,.
\end{equation}
Here $\tau_n$ is the average return time for the random walk to hit either edge
of the interval in the $n^{\rm th}$ excursion, conditioned on the walk
hitting either edge before it starves, while the term $\mathcal{S}$ accounts
for the final excursion in which the walk starves.  By definition
\begin{equation}
\label{tau-n}
\tau_n = \frac{\int_0^\mathcal{S} dt\,\, t \,\,F_n(t) }
{\int_0^\mathcal{S} dt\,  F_n(t) }= \frac{\int_0^\mathcal{S} dt\,\, t \,\,F_n(t) }
{\mathcal{F}_n } \,,
\end{equation}
with $\mathcal{F}_n=\int_0^\mathcal{S} dt\, F_k(t)$.  The calculation of
$\tau_n$ and subsequently $\tau$ are presented in \ref{tau} and the final
result is
\begin{equation}
  \label{tau-av}
  \tau\simeq 3.26786\,\,\mathcal{S}\qquad \qquad {\cal S} \gg 1.
\end{equation}

Thus the average lifetime of a starving random walk also scales linearly in
$\mathcal{S}$.  Roughly speaking, the same amount of time is spent in carving
out a desert to reach the critical length, then in enlarging it always via
the same edge, before finally attempting (and failing) to cross this desert.
The underlying distributions of visited sites and lifetimes are visually
similar and are characterized by a well-defined peak near the average value
and an exponential large-argument tail.

\section{Two Dimensions}
\label{sec:2d}

\begin{figure}[ht]
\centerline{\subfigure[]{\includegraphics[width=0.425\textwidth]{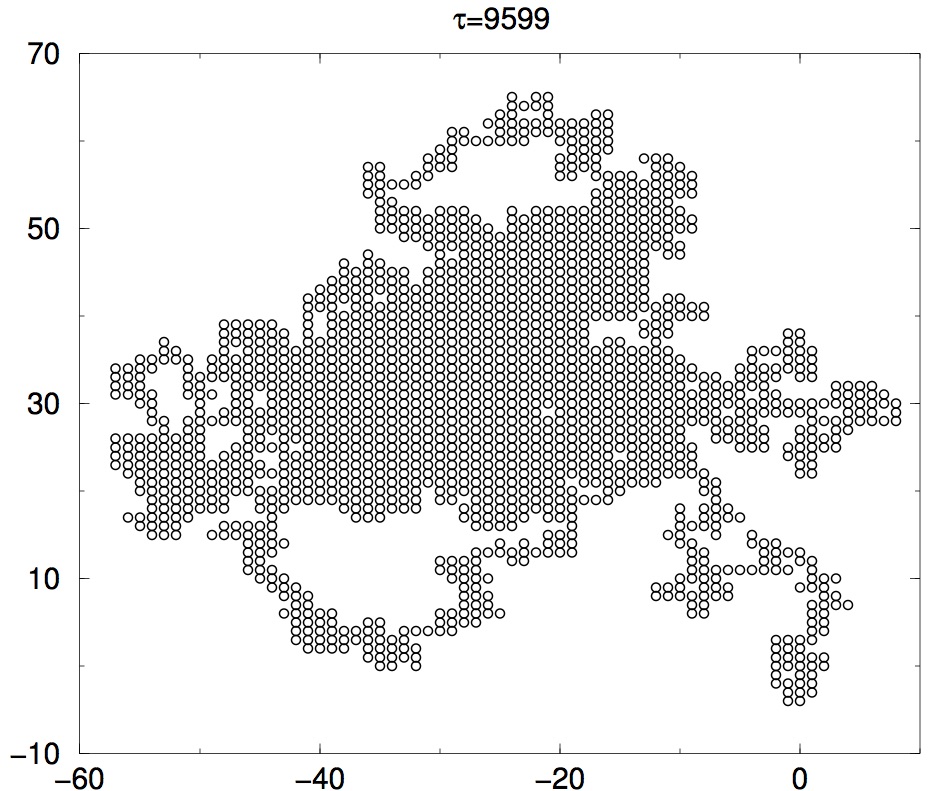}}\qquad
\subfigure[]{\includegraphics[width=0.425\textwidth]{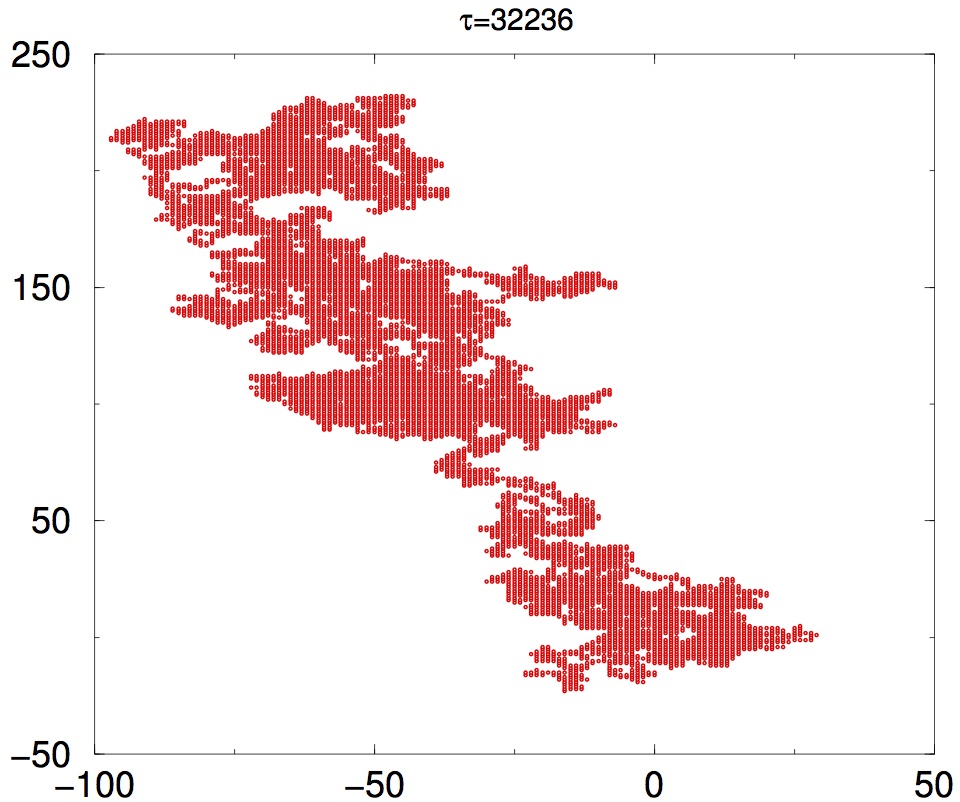}}}
\centerline{\subfigure[]{\includegraphics[width=0.415\textwidth]{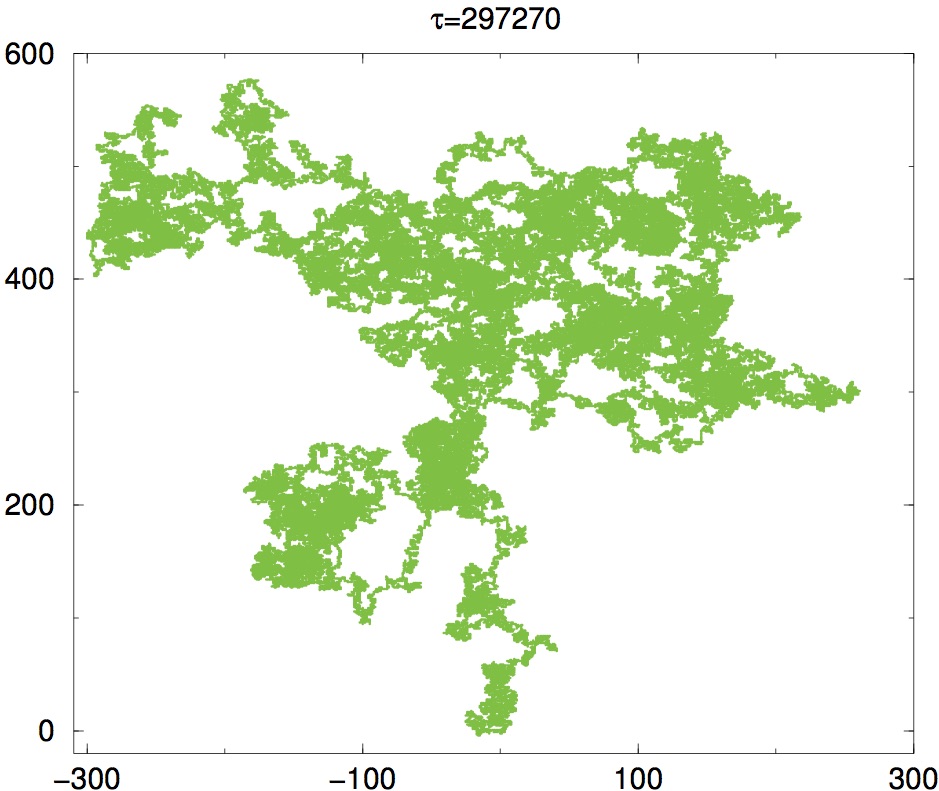}}\qquad
\subfigure[]{\includegraphics[width=0.425\textwidth]{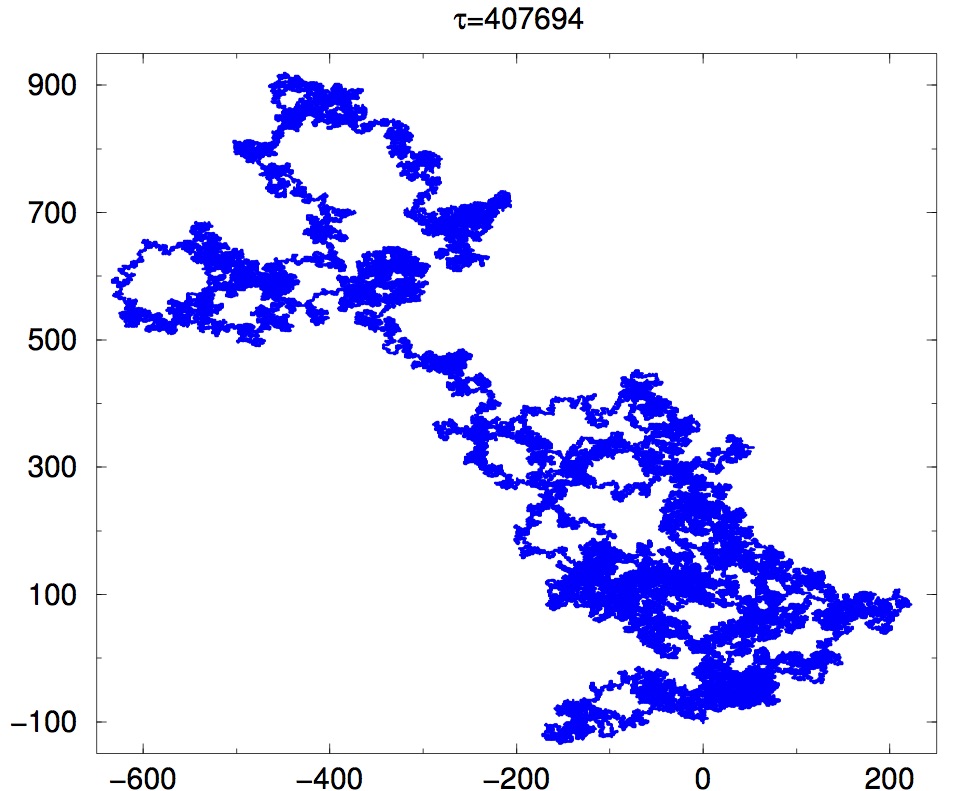}}}
\caption{Example random-walk trajectories for $\mathcal{S}=500$ with the
  lifetime $\tau$ of each walk indicated.  (a), (b): Short-lived walks.  (c),
  (d) Long-lived walks.}
\label{snapshots}
\end{figure}

Two dimensions is the ecologically relevant situation because of the obvious
connection to the movement of land foragers.  This case is also theoretically
interesting because $d=2$ is the critical dimension between recurrence and
transience of random walks~\cite{F68,W94,R01}.  Four sample trajectories of
starving random walks at the instant of starvation are given in
Fig.~\ref{snapshots} for metabolic capacity $\mathcal{S}=500$ to show the
types of trajectories that arise.  A walk survives if it never makes an
excursion of more than $\mathcal{S}$ steps anytime in its past history
without encountering food.  Thus the effect of starvation is to progressively
bias the ensemble of random walks as time increases.  For fixed
$\mathcal{S}$, a short-lived trajectory is relatively compact so that a
desert is quickly carved out within which the walk starves.  Conversely, a
long-lived walk typically has a ramified trajectory to ensure that the walk
remains close to food-containing sites and thus is unlikely to starve.

The starvation constraint on the trajectories is highly non-trivial and we
have been unable to solve the problem by exact methods.  Thus we turn to
simulations and crude approximations to elucidate some basic properties of
starving random walks in two dimensions.

\subsection{Simulation results}
\label{num2D}

Our simulations are typically based on $10^6$ realizations of nearest-neighbor
random walks that wander until they starve, for metabolic capacities
$\mathcal{S}$ between 10 and 2000.  For each walk, we record the time at the
instant of starvation, the position of the walk, and the number of distinct
sites visited (equal to the amount of food that the walk has eaten).  From
these data, we reconstruct the underlying distributions of these quantities.

\begin{figure}[ht]
\centerline{\subfigure[]{\includegraphics[width=0.45\textwidth]{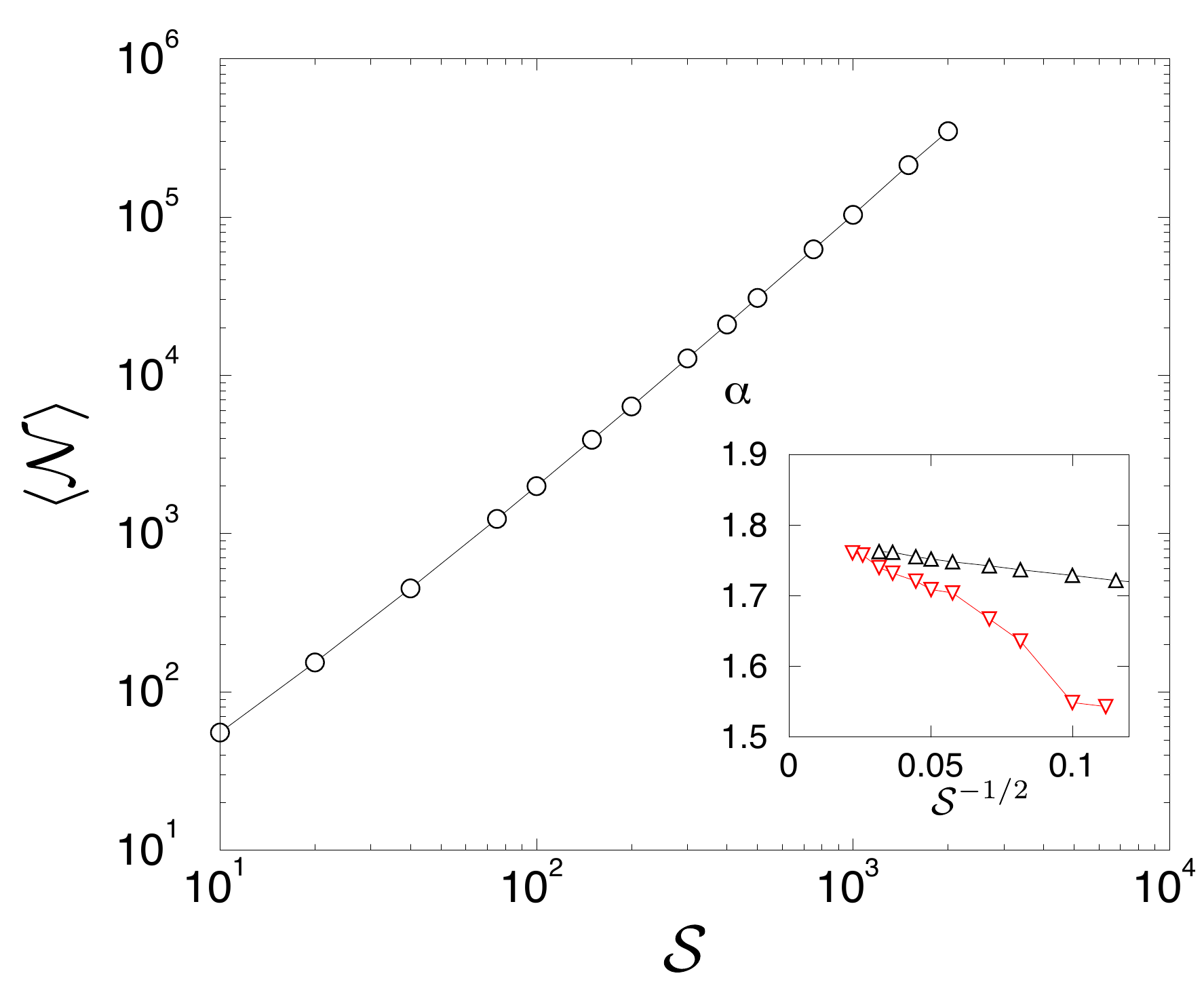}}\qquad
\subfigure[]{\includegraphics[width=0.45\textwidth]{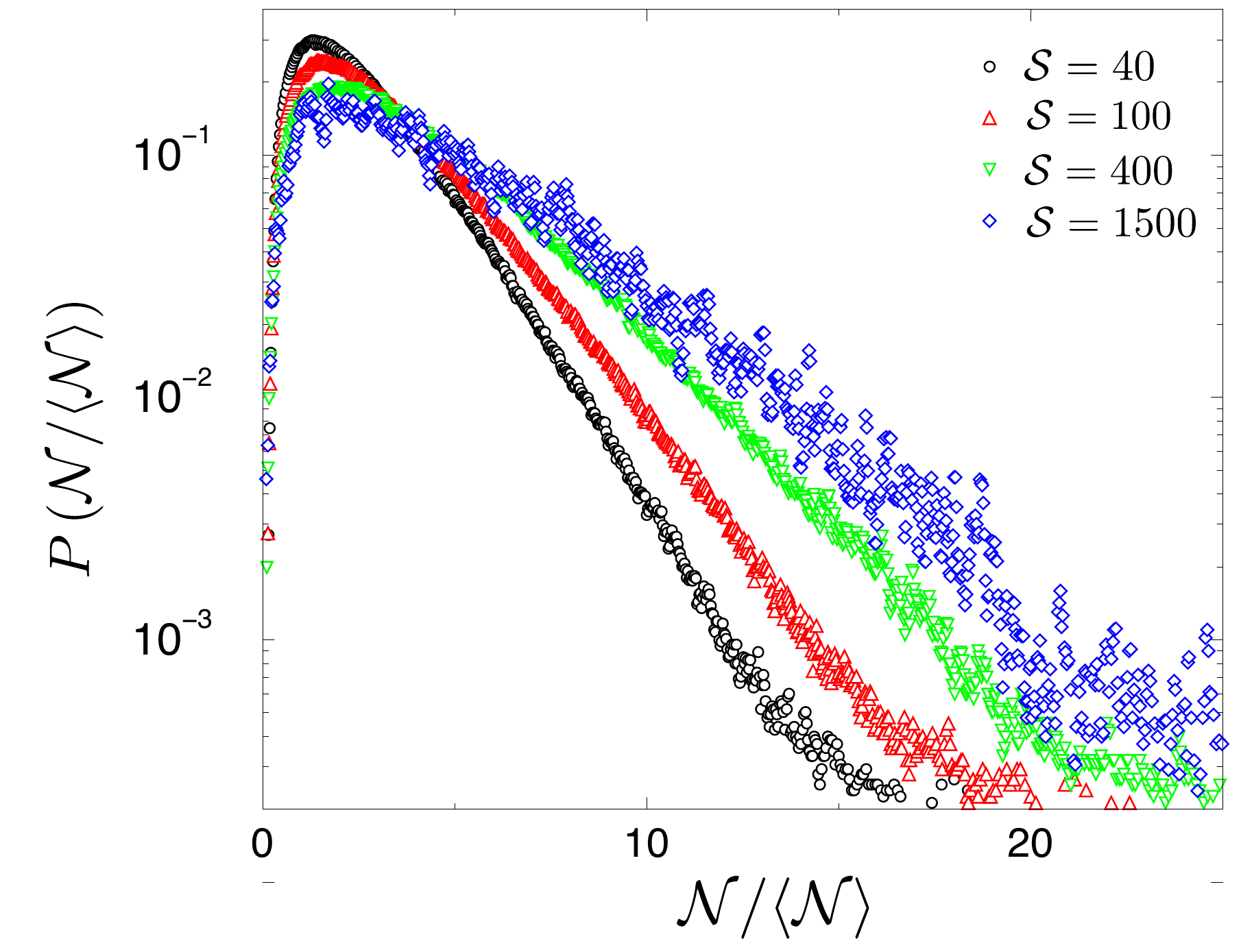}}}
\caption{(a) Average number of distinct sites visited for $10^6$ realizations
  of starving random walks at the starvation time as a function of
  $\mathcal{S}$.  The curve through the points is a guide for the eye.  The
  inset shows the local exponent ($\bigtriangleup$) and the running exponent
  ($\textcolor{red}{\nabla}$).  (b) The scaled distribution of number of
  distinct sites visited for 4 representative value of $\mathcal{S}$.  The
  data have been averaged over a 15-point range and only every fifth data
  point is displayed.  All data are based on $10^6$ for each value of
  $\mathcal{S}$.}
\label{s-av-2d}  
\end{figure} 

A plot of the average number of distinct sites visited at starvation
(Fig.~\ref{s-av-2d}) is suggestive of the algebraic behavior
$\langle \mathcal{N} \rangle \sim \mathcal{S}^\alpha$.  While a naive linear
fit to the data on a double logarithmic scale gives $\alpha\approx 1.66$,
there is a systematic upward curvature in the data of
$\ln\langle \mathcal{N} \rangle$ versus $\ln\mathcal{S}$.  We account for
this curvature by defining two types of $\mathcal{S}$-dependent exponents:
(a) a local exponent that is obtained from the slopes of four adjacent data
points in a window that is moved to progressively larger $\mathcal{S}$, and
(b) a running exponent that is obtained by successively deleting the first
point, the first two points, the first three points, etc.\ in the data.
These two $\mathcal{S}$-dependent exponents both appear to extrapolate to the
common value $\alpha \approx 1.8$ (Fig.~\ref{s-av-2d}(a) inset).  (For the
average lifetime and the root-mean-square displacement, we use only the
running exponent because it has smaller fluctuations.)~ The closeness of this
exponent to 2 and the fact that $d=2$ is the critical dimension of random
walks may indicative that the true dependence of
$\langle \mathcal{N} \rangle$ on $\mathcal{S}$ is quadratic but modified by a
logarithmic correction.

Perhaps the most surprising aspect of this data is that the distribution of
distinct sites visited at the instant of starvation does not satisfy
single-parameter scaling, as initially reported in~\cite{BR14}.  This is in
stark contrast to the cases of $d=1$ and $d=3$, where this same distribution
does obey single-parameter scaling.  This lack of scaling indicates that two
dimensions represents a unique situation.

\begin{figure}[ht]
\centerline{\subfigure[]{\includegraphics[width=0.43\textwidth]{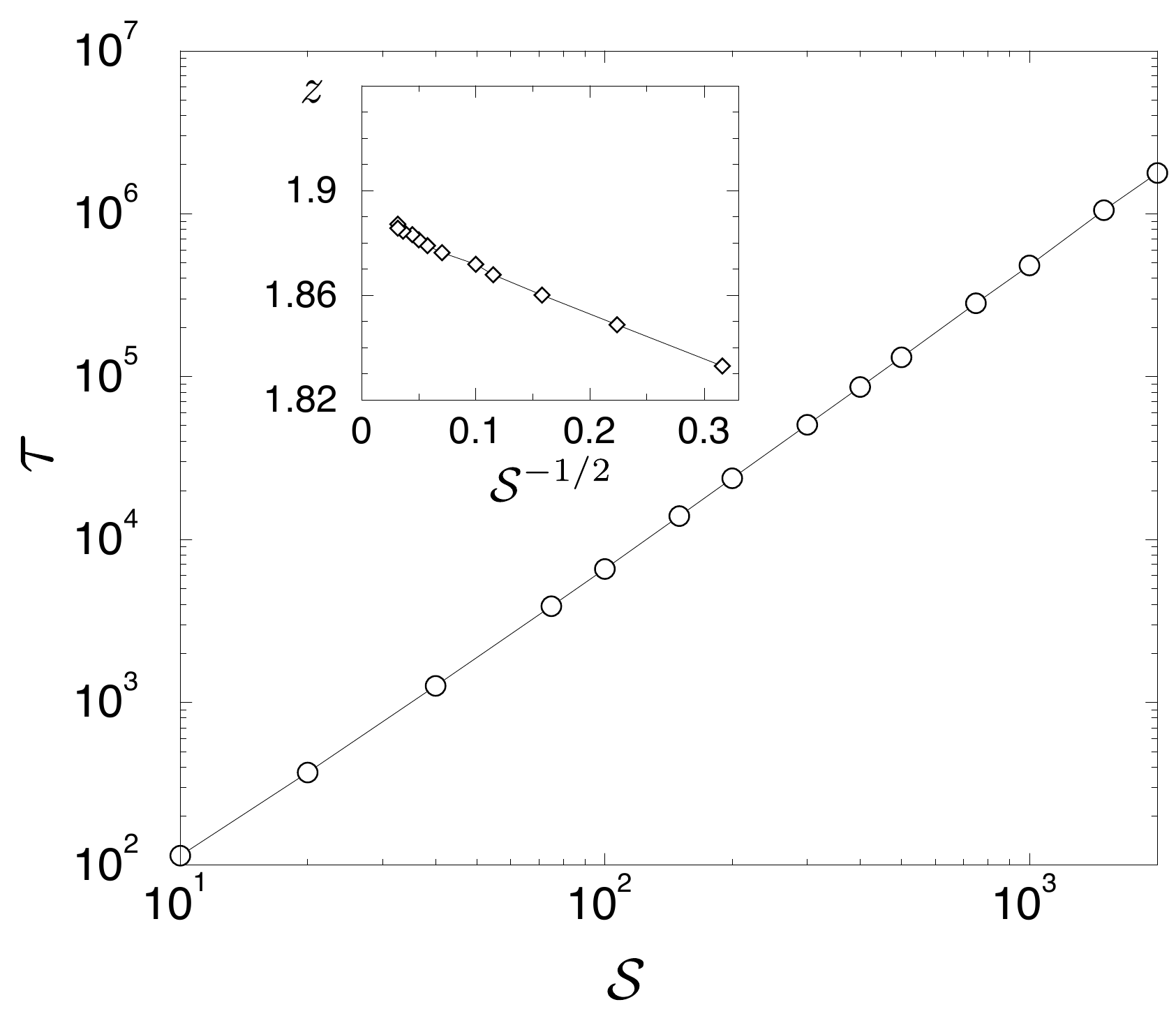}}\qquad
\subfigure[]{\includegraphics[width=0.45\textwidth]{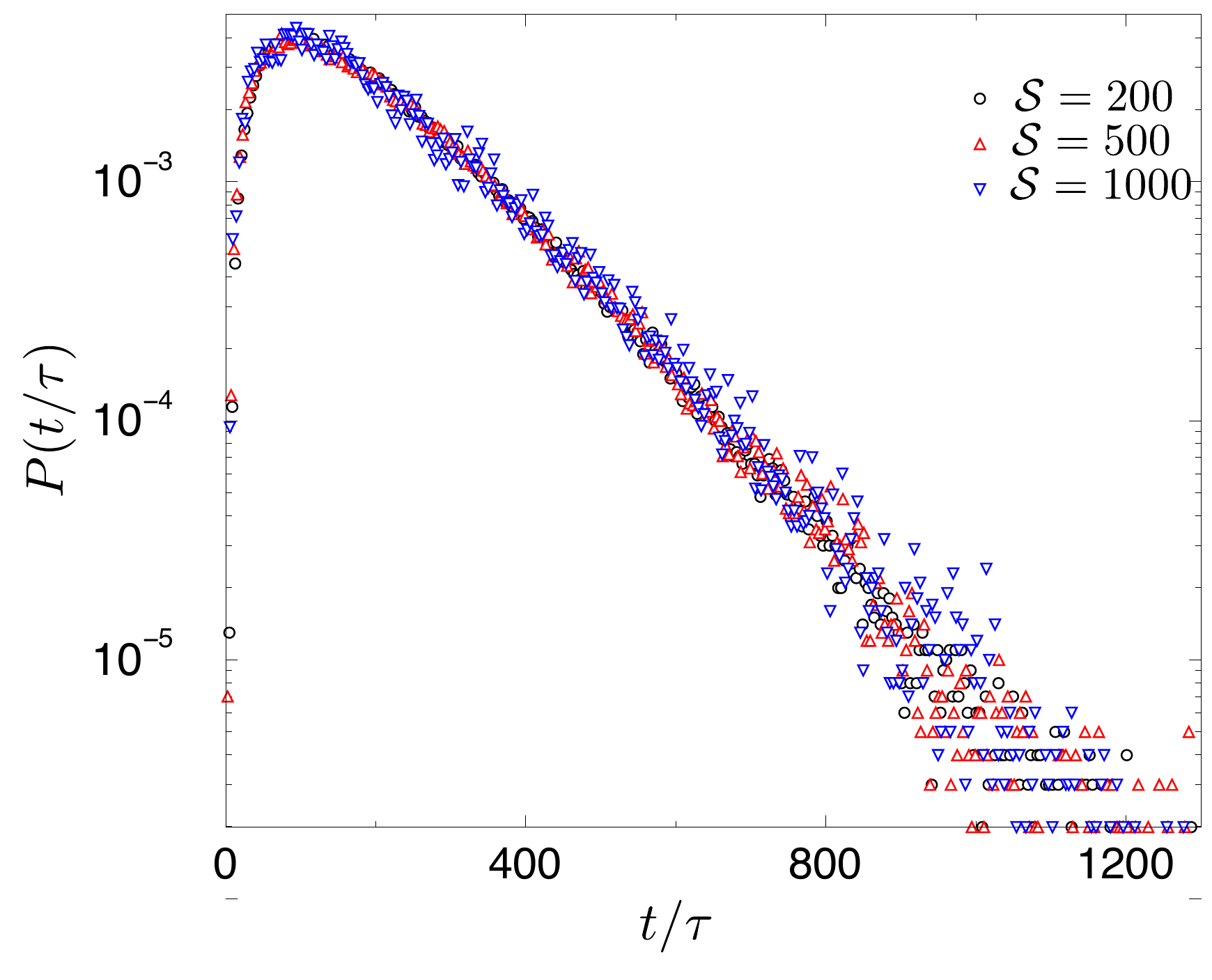}}}
\caption{(a) Average lifetime of starving random walks in $d=2$ as a function
  of $\mathcal{S}$.  The inset shows the running exponent.  (b) The scaled
  distribution of lifetimes $P(t/\tau)$ for 3 representative values of
  $\mathcal{S}$.  Data are based on $10^6$ realizations for each
  $\mathcal{S}$.}
\label{t-av-2d}  
\end{figure} 

The dependence of the average walk lifetime $\tau$ on $\mathcal{S}$ is
shown in Fig.~\ref{t-av-2d}.  Again, there is a small and systematic upward
curvature of the data of $\tau$ versus $\mathcal{S}$ on a double logarithmic
scale.  Thus we employ the same analysis as that used for the number of
distinct sites visited, to estimate that the asymptotic exponent for the
lifetime extrapolates to a value near 1.9 (inset to Fig.~\ref{t-av-2d}(a))
The closeness of this exponent estimate to 2 is again suggestive that perhaps
the dependence of $\tau$ on $\mathcal{S}$ may be modified by a logarithmic
correction.  The distribution of lifetimes for different values of
$\mathcal{S}$ collapse onto a single universal curve, with an exponential
long-time decay, when each distribution is scaled by the average lifetime
$\tau$ (Fig.~\ref{t-av-2d}(b)).  This contrasts with the lack of scaling
displayed by the distribution of visited sites.

\begin{figure}[ht]
\centerline{\subfigure[]{\includegraphics[width=0.43\textwidth]{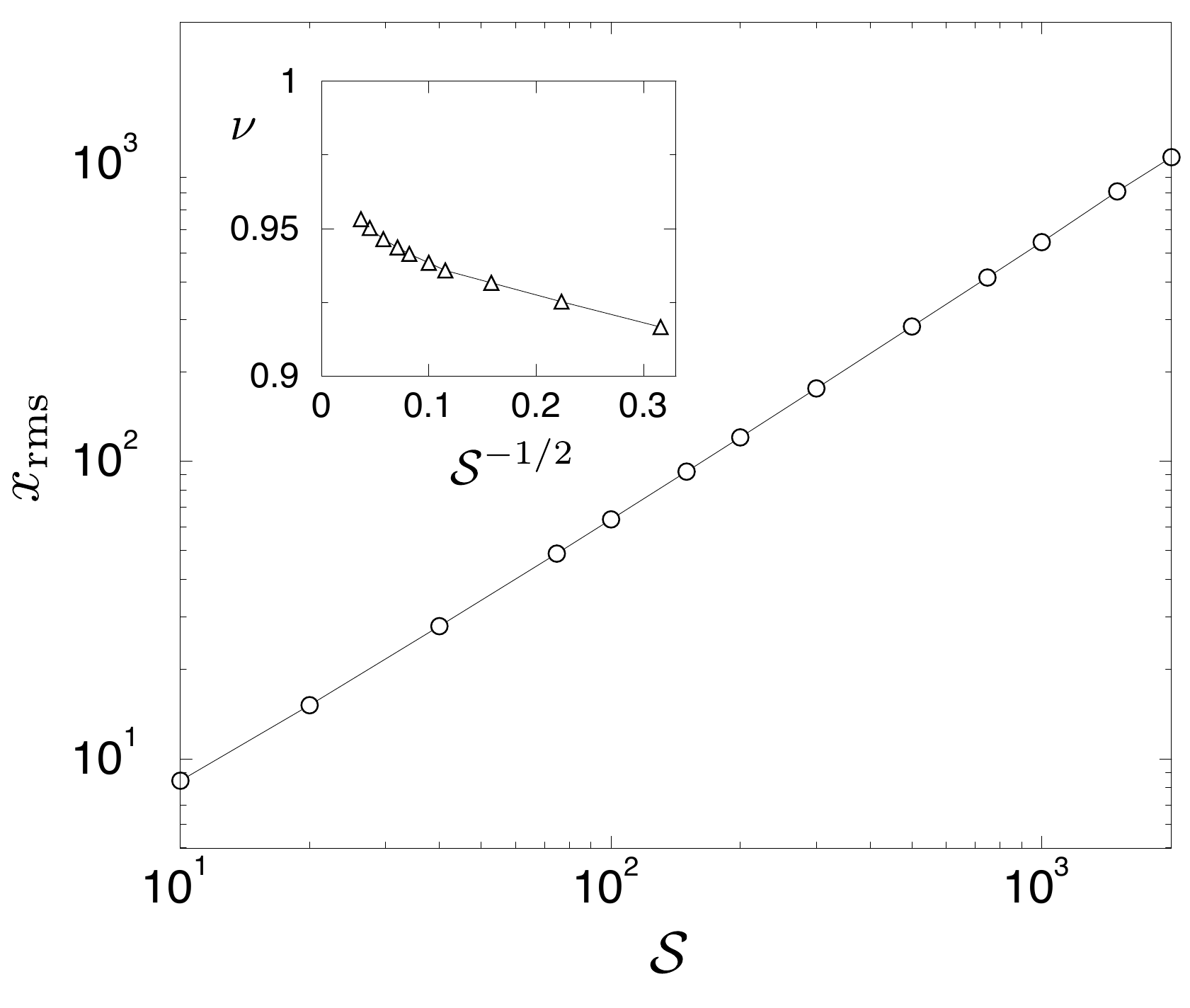}}\qquad
\subfigure[]{\includegraphics[width=0.45\textwidth]{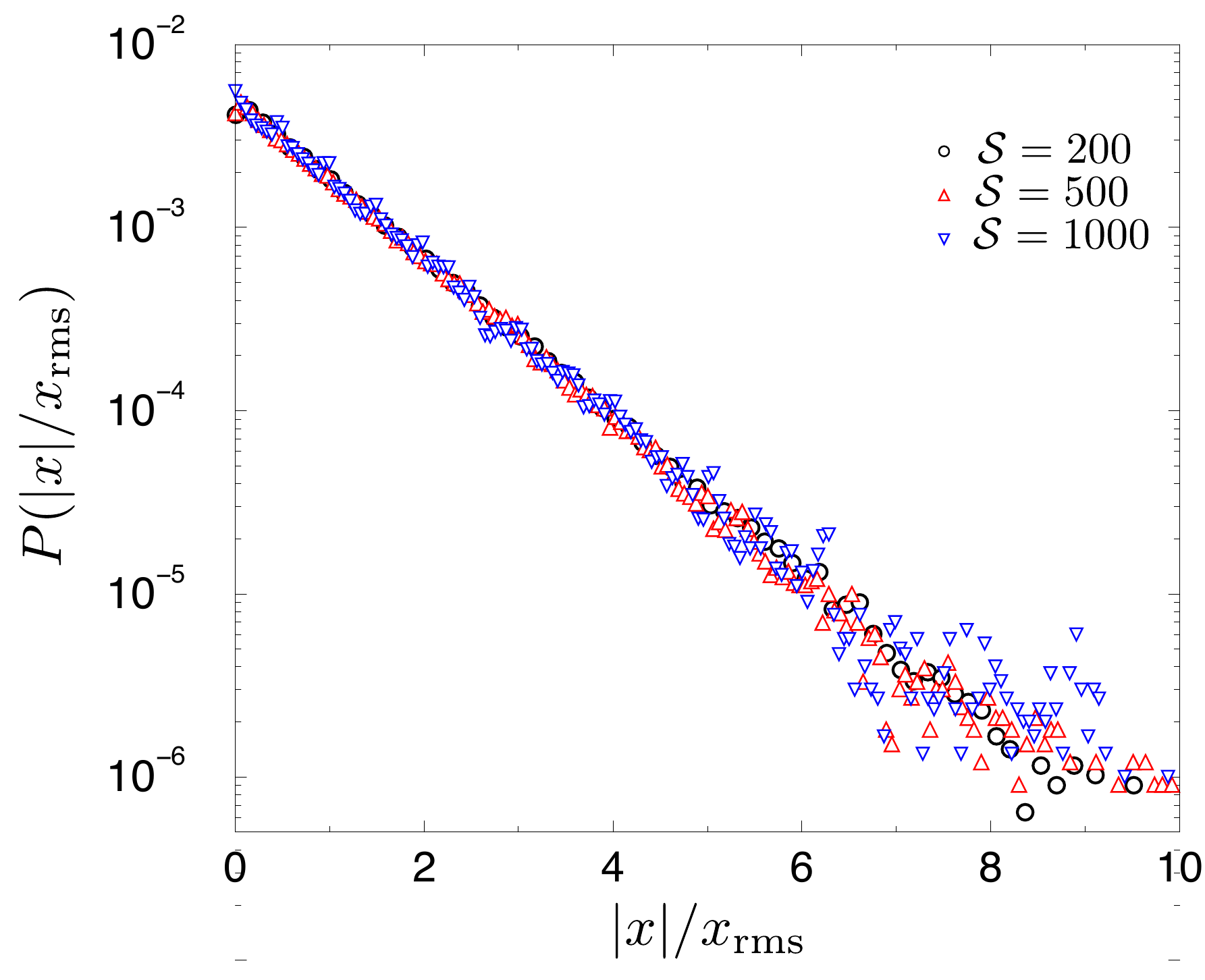}}}
\caption{(a) Average rms displacement of a single coordinate for starving
  random walks at the instant of starvation in $d=2$ as a function of
  $\mathcal{S}$.  The inset shows the running exponents in the main plot.  (b)
  Scaled distribution of this rms displacement for 3 representative values of
  $\mathcal{S}$.  Data are based on $10^6$ for each value of $\mathcal{S}$.}
\label{avr}  
\end{figure} 

Finally, we study the root-mean-square displacement of a single Cartesian
coordinate $x_{\rm rms}\equiv \sqrt{\langle x^2\rangle}$ and its distribution
at the instant of starvation.  As shown in Fig.~\ref{avr}(a), it appears that
$x_{\rm rms}\sim \mathcal{S}^{\nu}$, with $\nu$ close to 1.  This dependence
accords with naive expectations based on the dependence of $\tau$ on
$\mathcal{S}$.  Namely, if $x_{\rm rms}\sim \mathcal{S}$ and
$\tau\sim \mathcal{S}^2$, and furthermore, the distribution of lifetimes is
sharply peaked about its most probable value, then we should expect that
$x_{\rm rms}\sim\sqrt{\tau}$, which is the same scaling of rms displacement
versus time as an unrestricted random walk.  Finally, Fig.~\ref{avr}(b) shows
the scaled distribution of the absolute value of a single coordinate at the
instant of starvation.  The distributions for different ${\cal S}$ all
collapse onto a universal curve that decays exponentially, in contrast to the
Gaussian decay of a pure random walk.  This distinction stems from the
difference between the ensemble of trajectories of a starving random walk
and that of an unconstrained random walk. Indeed, the long-lived trajectories
for a starving random walk are more ramified and more spatially extended than
for a pure random walk, as the walk needs to remain close to resources to
survive.

\subsection{Circular approximation: heuristic approach}

Because of the typically labyrinthine desert geometry in two dimensions, it
does not seem feasible to determine the properties of two-dimensional
starving random walks exactly.  Thus we develop an approximation that is
based on the assumption that the desert remains circular at all times.  While
obviously crude (compare with Fig.~\ref{snapshots}), this approximation
provides a lower bound for the average lifetime at starvation because a circular
desert is the most unfavorable geometry for the walk to survive.

\begin{figure}[ht]
\centering
\includegraphics[width=0.7\textwidth]{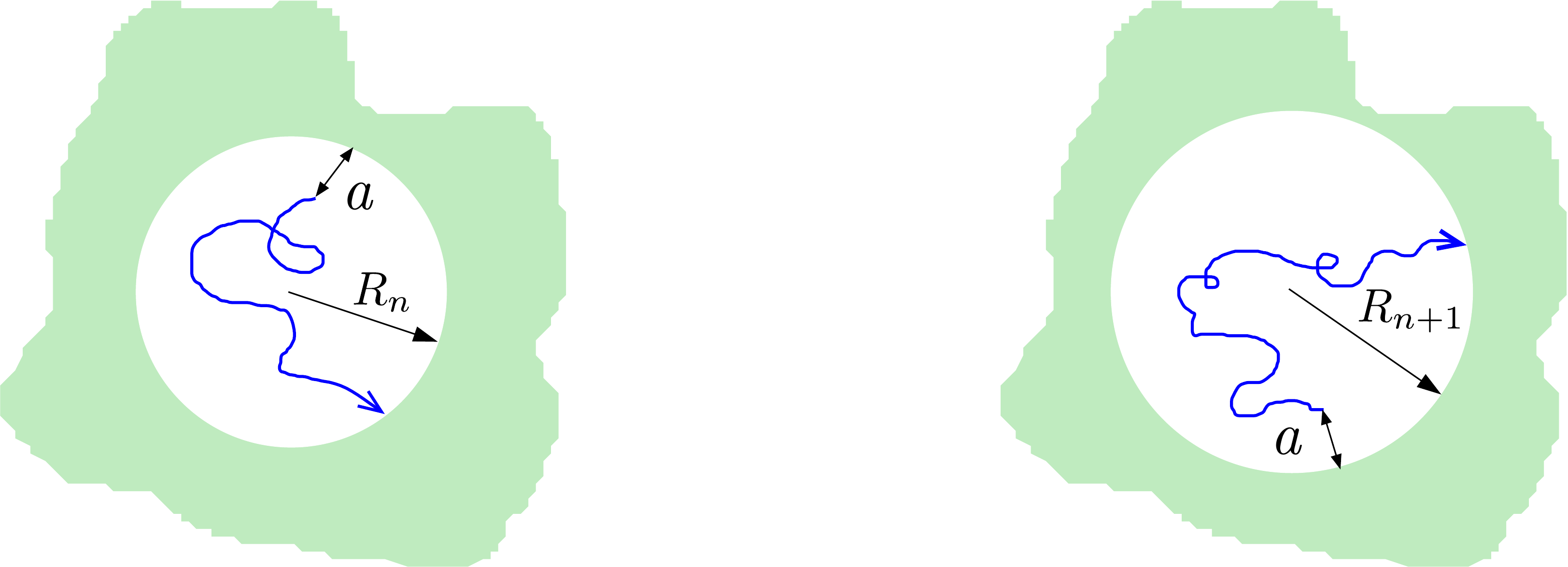}
\caption{Schematic illustration of the evolution of the desert in the
  circular approximation.  When the random walk hits the boundary, it is
  restarted a distance $a$ from the newly enlarged boundary.}
\label{circular}  
\end{figure}

We now adapt the heuristic argument given in~\ref{heuristic} for one
dimension to this circular approximation. Each time the walk hits the edge
of the desert, it eats one unit of food, which we take to be a small patch of
area $a^2$, with $a$ the lattice spacing.  After each meal, the area of the
desert increases by $a^2$ so that its radius after $n$ meals is
$R_n=\sqrt{n a^2/\pi}$.  Moreover, we assume that after reaching the desert
boundary, the walk starts a distance $a$ from the enlarged boundary to
begin another excursion (Fig.~\ref{circular}), as on the lattice.

Following Sec.~\ref{heuristic}, we again decompose a typical trajectory into
three phases:
\begin{enumerate}
\item The walk carves out a circular desert of a ``dangerous'' radius $R_c$.
\item After $R_c$ is reached, the walk makes $M$ excursions into the desert
  without starving.
\item The walk wanders too deeply into the desert and starves.
\end{enumerate}
The desert radius reaches a dangerous value $R_c$ when the average time to go
from the edge to the center is roughly one-half of the metabolic capacity
$\mathcal{S}$.  This typically corresponds to the average conditional time to
reach a small disk in the center of the desert without having touched its
edge, at the radius $R_c$, starting from a radius $R_c-a$~\cite{R01}.  This
gives the criterion
\begin{equation*}
\label{tr}
\frac{R_c^2}{4D} \simeq \frac{{\cal S}}{2}\,,
\end{equation*}
from which $R_c= \sqrt{2D\mathcal{S}}$.

We now estimate the time to reach this critical radius.  When the desert is
subcritical, a forager typically survives an excursion no matter where it
exits the desert.  For a walk that starts at radius $R_n\!-\!a$, the time
to reach the edge of the desert is $t_n\sim{aR_n}/{2D}$.  The number of meals
$n_c$ that the forager must eat for the desert to reach the critical radius
is determined from $\pi R_c^2= n_ca^2$, or $n_c= 2\pi D\mathcal{S}/a^2$.
Thus the duration of phase (i) is the sum of the excursion times until the
critical radius is reached:
\begin{align}
T_1&=\sum_{n<n_c} t_n = \sum_{n<n_c} \frac{aR_n}{2D} 
\simeq\frac{a}{3D}\sqrt{\frac{a^2}{\pi}}\,n_c^{3/2}
\sim \frac{D^{1/2}}{a}\,\mathcal{S}^{3/2}\,.
\end{align}

As we will show, the duration of phases (ii) and (iii) are negligible
compared to phase (i).  Phase (iii) has duration $\mathcal{S}$ by definition.
For phase (ii), if we know the typical number $M$ of times that the walk
returns to the edge of the desert of initial radius $R_c$ before starving,
then its duration is given by
\begin{equation}\label{T2}
T_2=\sum_{n=n_c}^{n_c+M} R_n \simeq \int_{n_c}^{n_c+M}\!\!\! \sqrt{n}\,dn \,\, \propto \,\, (n_c\!+\!M)^{3/2}-n_c^{3/2}\,.
\end{equation} 
While we are unable to determine $M$, we can provide an upper bound that we
write as $M_+$.  Suppose that the desert radius remains fixed at $R_c$ each
time the walk returns to the boundary.  The number of returns $M_+$ in this
case is strictly greater than $M$.  To estimate $M_+$, we need the
probability $\mathcal{E}$ that the walk does not starve in this fixed-size
desert between two consecutive returns to the edge of the desert.  We
estimate $\mathcal{E}$ as the probability that the walk does not reach the
core of the desert, which we take as a small disk of radius $b$ that is much
less than $R_c$.  This quantity is just the splitting probability for a
walk that starts at radius $R_c-a$ to return to radius $R_c$ before
reaching radius $b$.  For unconstrained diffusion, this splitting probability
is given by~\cite{R01}
\begin{equation}
\mathcal{E}=\frac{\ln\left[(R_c-a)/b\right]}{\ln \left({R_c}/{b}\right)} \simeq 1-\frac{C}{\sqrt{\mathcal{S}} \ln\mathcal{S}} \quad {\rm for}~{\cal S} \gg 1.
\end{equation}
with $C$ a constant of order 1 as long as $b\ll R_c$.  The probability that
the walk returns $k$ times to the outer boundary $R_c$ before starving is
$\Pi_k=\mathcal{E}^k \, (1-\mathcal{E})$.  Hence the average number $M_+$
of returns to this boundary is
\begin{equation}
M_+=\sum\limits_{k=0}^{\infty} k \, \Pi_k =\frac{\mathcal{E}}{1-\mathcal{E}}  \; \propto \sqrt{\mathcal{S}} \ln \mathcal{S} \,.
\end{equation}
Since we previously showed that $n_c\propto \mathcal{S}$, we have
$M < M_+ \ll n_c$.  Thus Eq.~\eqref{T2} yields
\begin{equation}
T_2 \propto M \sqrt{n_c} < \mathcal{S} \ln \mathcal{S}\,.
\end{equation}
Thus phase (ii) has negligible duration compared to that of phase (i), so
that the average lifetime $\tau$ of the walk in the circular approximation
scales as $\mathcal{S}^{3/2}$. Similarly, the number $M$ of returns to the
edge of the desert during phase (ii) is negligible compared to the number
$n_c$ of such returns during phase (i), so that $\langle {\cal N} \rangle$,
the average number of distinct sites visited in the circular approximation
scales as ${\cal S}$.

\subsection{Circular approximation: microscopic approach}

We now determine the distribution $V(\mathcal{N})$ of the number of distinct
sites visited at starvation in the circular approximation and then extract
the average lifetime and the average number of distinct sites visited.  As in one
dimension, we write $V(\mathcal{N})$ as
\begin{equation}
V(\mathcal{N})= \mathcal{F}_2 \,\mathcal{F}_3\, \mathcal{F}_4 \ldots \mathcal{F}_\mathcal{N} (1-\mathcal{F}_{\mathcal{N}+1})\,, \nonumber
\end{equation}
where again $\mathcal{F}_k=\int_0^\mathcal{S} dt\, F_k(t)$, with $F_k(t)$ the
first-passage probability at time $t$ to any point of the circular boundary
of radius $R_k$ when the walk starts a distance $a$ from this boundary.
Following a similar approach as in the case of one dimension (see~\ref{P2D}),
we obtain the following expression for the distribution of visited sites at
the instant of starvation\footnote{Equation~\eqref{Pcirc} corrects errors in the corresponding
  formula in~\cite{BR14}.}
\begin{equation} 
\label{Pcirc}
V(\theta,\mathcal{S}) \simeq \frac{2 \pi}{a} \sqrt{\frac{D
    \mathcal{S}}{\theta}} 
\sum\limits_{k=1}^{\infty} e^{-{j_k^2}/{\theta}}  \exp \left\{\!- \frac{4 \pi}{a} \sqrt{D\mathcal{S} \theta} \sum\limits_{m=1}^{\infty} \left[ e^{-{j_m^2}/{\theta}}  -\sqrt{\frac{\pi}{\theta}}\, j_m \, {\rm erfc}\left(\frac{j_m}{\sqrt{\theta}}\right) \right]\! \right\} \,
\end{equation}
where $\theta\equiv \mathcal{N} a^2/(\pi D \mathcal{S})$.  Because of the
simultaneous appearance of $\mathcal{S}$ and $\theta$, this distribution of
the number of distinct sites visited does not obey single-parameter scaling, as
mentioned in Sec.~\ref{num2D}.

A numerical evaluation of \eqref{Pcirc} is not straightforward because the
sums converge slowly and it appears that average number of distinct sites
visited grows as
\begin{subequations}
\begin{equation}
\label{Nav2D}
\langle \mathcal{N} \rangle \sim \mathcal{S}^{\beta}
\end{equation}
with $\beta \simeq 0.9$.  This value roughly accords with the scaling that
was obtained from the heuristic circular approximation.  Using the circular
approximation, we may also deduce the average lifetime of the walk
(\ref{tau2D}).  Subject to the same convergence issues as in the number of
distinct sites visited, the lifetime appears to grow as
\begin{equation}
\label{tauav2D}
\tau \sim \mathcal{S}^{\gamma}
\end{equation}
\end{subequations}
with $\gamma \simeq 1.4$.  This exponent estimate is also close to that
obtained in the heuristic circular approximation.

The exponents in Eqs.~\eqref{Nav2D} and \eqref{tauav2D} are substantially
smaller than the corresponding values obtained by numerical simulations,
$\beta\approx 1.8$ and $\gamma\approx 1.9$.  Thus the circular approximation
is a relatively weak lower bound for the starving random walk on the lattice.
This disparity is not unexpected because the true shape of the desert is
generally quite ramified (Fig.~\ref{snapshots}).  Such a desert shape allows
the walk to eat much more often than in the circular approximation.  In
spite of the imprecise numerical results, the circular approximation gives a
lower bound for $\langle \mathcal{N} \rangle$ and $\tau$ in two dimensions,
where the complex desert shape seems to render an exact calculation
unfeasible.  The circular approximation also captures the two-parameter
scaling for the distribution of the number of distinct sites visited, a
property that does not exist in one dimension.

\section{Infinite-Dimensional Limit}
\label{sec:infd}

For large spatial dimension $d$, an $N$-step random walk visits $A\,N$
distinct sites on average~\cite{F68,W94}, with $A$ a constant that approach 1
as $d\to\infty$.  Thus the depletion of the environment plays an
insignificant role as the walk typically visits new (food-containing sites)
at each step.  We therefore investigate the starvation dynamics for large $d$
under the assumption that the walk hits a previously visited or a
previously unvisited site at each step with time-independent probabilities
$\lambda$ and $1-\lambda$ respectively.  For a hypercubic lattice in $d\gg 1$
dimensions, a naive estimate for the probability $\lambda$ is just the
backtracking probability $1/(2d)$.  The relevant point is that $\lambda$ is
very small in the high-dimensional limit.

\begin{figure}[ht]
\centerline{\includegraphics[width=0.6\textwidth]{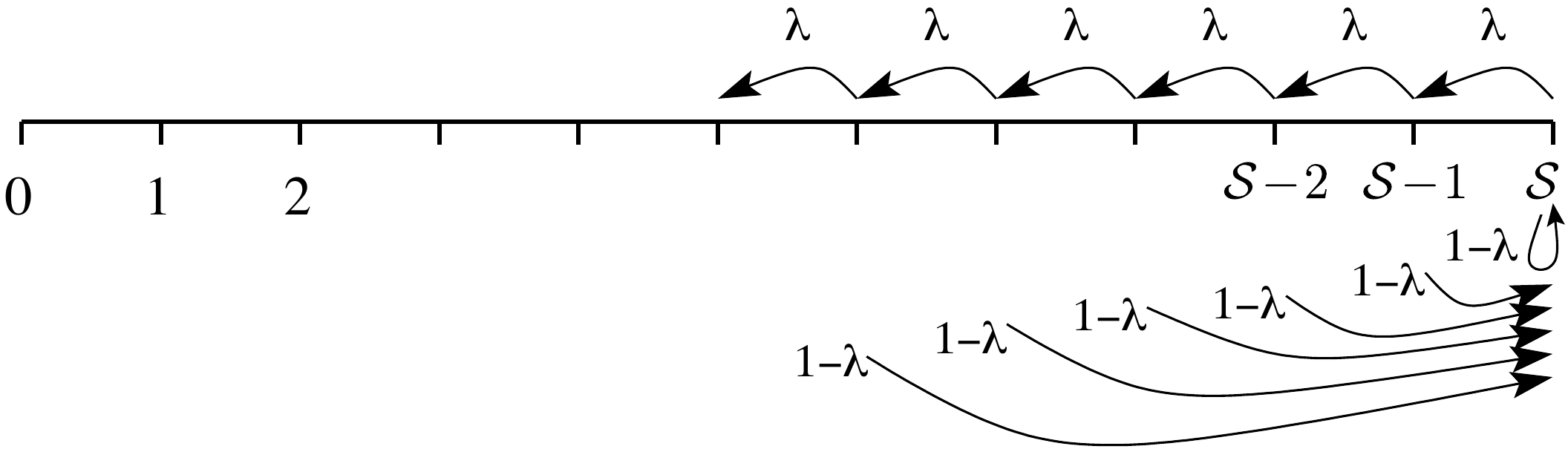}}
\caption{Evolution of a high-dimensional random walk in ``starvation space''.
  A walk at position $n$ can survive $n$ additional steps without
  encountering food.}
\label{line}  
\end{figure} 

Under the assumption that visits to full or empty sites are independent
events with the probabilities given above, the time until the walk starves
undergoes a stochastic process in a one-dimensional ``starvation space''---
an interval of length $\mathcal{S}$.  When the origin is reached in this
starvation space, it corresponds to the starvation of the walk on the
lattice (Fig.~\ref{line}).  A walk at site $n$ in starvation space can
wander $n$ additional steps on the lattice without encountering food before
it starves.  When the walk hits a previously visited (empty) site in
physical space, the time to starvation decreases by one.  This corresponds to
a hop to the left (with probability $\lambda$) in starvation space.
Conversely, when the random walk hits a new (food-containing) site, it can
wander an additional $\mathcal{S}$ steps on the lattice until starvation
occurs. This corresponds to an immediate hop to site $\mathcal{S}$ in
starvation space with probability $1-\lambda$.

Based on this picture, we compute $t_n$, the average time until the walk
starves when starting from position $n$ in starvation space.  These average
starvation times satisfy the backward equations~\cite{R01}:
\begin{align}
\begin{split}
&t_{1} = (1-\lambda)t_\mathcal{S}+1\,, \\
&t_n = \lambda t_{n-1} + (1-\lambda) t_\mathcal{S}+1\hskip 1in
2\leq n\leq \mathcal{S}\,.
\end{split}
\end{align}
In the equation for $t_n$, the first term accounts for hopping to a
previously-visited site, in which case the walk can wander $n-1$ additional
steps before starving.  The second term accounts for hopping to a new
food-containing site, in which case the walk can wander $\mathcal{S}$ more
steps before starving.  The last term accounts for the time elapsed for a
single step.  We use the first equation to eliminate $t_1$ in the recursion
for $t_2$, and thereby determine $t_2$ in terms of $t_\mathcal{S}$.
Repeating this elimination gives each $t_n$ in terms of $t_\mathcal{S}$, and
ultimately the closed equation for $t_\mathcal{S}$
\begin{equation}
\label{tS}
t_\mathcal{S}= \frac{1}{\lambda^\mathcal{S}}\,\Big(\frac{1-\lambda^\mathcal{S}}{1-\lambda}\Big)~.
\end{equation}
In the (unrealistic) limit of $\lambda\to 1$, the average starvation time
approaches $\mathcal{S}$, while, for the relevant high-dimensional limit
where $\lambda\to 0$, the average starvation time grows exponentially with
$\mathcal{S}$.

We may also solve for the probability that the walk starves at a given time.
Let $Q_n(t)$ be the probability that a walk at time $t$ can wander without
food for an additional $n$ steps before starving.  This probability obeys the
recursion
\begin{subequations}
\begin{align}
\begin{split}
\label{Qn}
&F(t) = \lambda Q_1(t-1)\,,\\
&Q_n(t) = \lambda Q_{n+1}(t-1)\hskip 2in  1\leq n\leq \mathcal{S}-1\,,\\
&Q_\mathcal{S}(t)=
(1\!-\!\lambda)\big[Q_\mathcal{S}(t\!-\!1)+Q_{\mathcal{S}-1}(t\!-\!1) +\ldots
+Q_1(t\!-\!1)\big]\,.
\end{split}
\end{align}
The first equation accounts for a walk that is one time step from starvation
and hops to an empty site, and thus gives the first-passage probability that
the walk dies at time $t$.  The second equation accounts for processes in
which a walk hops to an empty site and thus moves one time unit closer to
starvation.  The last equation accounts for events in which a walk hops from
any point in starvation space to a full site.  When this occurs, the walk
can survive $\mathcal{S}$ additional time units without eating before it
starves.

We impose the initial condition $Q_n(t\!=\!0)=\delta_{n,\mathcal{S}}$,
corresponding to the walk being fully sated at the outset.  To solve
Eqs.~\eqref{Qn}, we multiply each equation by $z^t$, and sum from $t=1$ to
$\infty$.  In terms of the the generating function
$Q_n(z)=\sum_{t\geq 0} Q_n(t) z^t$, we obtain the algebraic equations
\begin{align}
\begin{split}
&F(z) = \lambda z Q_1(z)\,, \\
&Q_n(z) = \lambda z Q_{n+1}(z)\hskip 1.8in 1\leq n\leq \mathcal{S}-1\,, \\
&Q_\mathcal{S}(z) = 1+z(1-\lambda)\sum_{1\leq k\leq \mathcal{S}}Q_n(z)\,.
\end{split}
\end{align}
\end{subequations}
Solving these equations, the generating function for the first-passage
probability is
\begin{equation}
F(z) = \frac{(z\lambda)^\mathcal{S}(1-\lambda z)}{1-z\big[1-(1-\lambda)(\lambda z)^\mathcal{S}\big]}~.
\end{equation}
Notice that $F(z\!=\!1)=1$, which means that the sum of the first-passage
probability over all times equals 1.  Thus the walk eventually must starve.
One can also readily verify that the average first-passage time given in
\eqref{tS} is reproduced by
\begin{equation}
\langle t\rangle = \sum t F(t) = z \frac{\partial F(z)}{\partial z}\bigg|_{z=1}\,.
\end{equation}
Because $F(z)$ has a simple pole at
$z=z_c\approx 1+(1-\lambda)\lambda^\mathcal{S}$ to lowest order, the
first-passage probability decays exponentially with time.  The characteristic
decay time of the first-passage probability is given by $z_c^{-1}$, which has
the same scaling as the average first-passage time given in Eq.~\eqref{tS}.

Finally, we determine the distribution of the number of sites visited by the
walk on the lattice.  The essential ingredient to determine this distribution
is the return probability $R_n$, namely, the probability for a walk that
starts from site $n$ in starvation space to reach $\mathcal{S}$ without first
hitting 0.  Each such return corresponds to the random walk visiting a new
site on the physical lattice without starving.  These return probabilities
satisfy the recursion
\begin{align}
\begin{split}
R_1&=(1-\lambda)\,, \\
R_n&= (1-\lambda) + \lambda R_{n-1}    \hskip 1in 2\leq n\leq \mathcal{S}\,.
\end{split}
\end{align}
In the equation for $R_n$, the first term accounts for the walk hitting a
previously unvisited site, in which case the walk is at site $\mathcal{S}$ in
starvation space, while the second term accounts for the walk hitting an
empty site, in which case the walk is at site $n-1$ in starvation space.

The solution to this recursion is simply $R_n=1-\lambda^n$.  From these
return probabilities, we now infer the distribution of distinct sites visited
at starvation.  For a walk that starts at site $\mathcal{S}$ in starvation
space, the probability that $\mathcal{N}$ distinct sites are visited on the
lattice before the walk starves is simply given by
\begin{equation}
  P(\mathcal{N})= R_\mathcal{S}^{\mathcal{N}}(1-R_\mathcal{S})\,.
\end{equation}
This equation expresses the fact that to visit $\mathcal{N}$ distinct sites
on the lattice, the walk must visit a full site $\mathcal{N}$ times, each
within $\mathcal{S}$ steps, before the walk staves.  From this result, the
average number of distinct sites visited before the walk starves is
\begin{equation}
\langle \mathcal{N}\rangle = \frac{R_\mathcal{S}}{1-R_\mathcal{S}}= \lambda^{-\mathcal{S}}(1-
\lambda^{\mathcal{S}})\,.
\end{equation}
As might be expected in this mean-field description, the average lifetime
given in Eq.~\eqref{tS} and the average number of distinct sites visited
before starving are the same up to an overall constant factor of $1-\lambda$,
and both grow exponentially with the metabolic capacity ${\cal S}$. These
long-lived trajectories in large spatial dimension arise from the low
probability to encounter a previously visited site.

\section{Summary and Discussion}
\label{sec:conc}

The starving random walk model represents a minimalist description for the
consumption of a depleting resource by an unaware forager.  While there is
much literature on optimizing the search for resources by a rational forager
that possesses complete knowledge of its environment (see,
e.g.,~\cite{C76,B91,OB90,KM01,ASD97,KR85,SK86,Vea96,LKW88}), the
complementary starving random walk model in which the forager has no propensity
to move toward food and no environmental knowledge seems mostly unexplored.
The fundamental parameters for a starving random walk are its metabolic
capacity $\mathcal{S}$, namely, the number of steps that the walk can wander
in a row without food before starving to death, and secondarily the spatial
dimension $d$.  The latter parameter enters because of the non-trivial
dimension dependence of visits to distinct sites for random walks.

We solved the starving random walk model in one dimension by probabilistic
methods to give the average lifetime of the walk, the average number of
distinct sites visited at the instant of starvation, and the distribution of the latter.
Qualitatively, a forager carves out a growing desert until the desert reaches
a dangerous critical length of the order of $\sqrt{D\mathcal{S}}$.  When this
length is reached, the walk can still survive if it returns to the same side
of the desert, but starves when it attempts to cross to the other side.  From
this picture, the average number of distinct sites visited at the instant of
starvation, which is the same as the total amount of consumed food, scales as
$\sqrt{\mathcal{S}}$, while the average lifetime scales as $\mathcal{S}$.

In the ecologically relevant case of two dimensions, a probabilistic solution
for the lifetime and number of distinct sites visited remains an open
challenge.  By numerical simulations, both the average number of distinct
sites visited and the average lifetime appear to scale algebraically with
$\mathcal{S}$, with an exponent close to 1.8 for the former quantity and
exponent close to 1.9 for the latter.  The spatial probability distribution
of starving random walks is exponential rather than the Gaussian for a pure
random walk.  This difference reflects the feature that a long-lived starving
random walk remains close to the source of food, which means that it should
explore a ramified and more spatially extended region than a pure random
walk.  We provided a (weak) lower bound for the average lifetime and number of
distinct sites visited within a circular approximation. The two-dimensional
case is particularly challenging, as the distribution of the distinct sites
visited does not obey single-parameter scaling~\cite{BR14} and the relation
between the geometry of the carved out desert and the lifetime of the walk
is not yet resolved.

In greater than two dimensions, the transience of the random walk means that
visits to new (i.e., food containing) sites occur at a fixed rate.  Thus a
starving random walk is much more long lived for $d>2$ than in $d\leq 2$.
Indeed, we previously found~\cite{BR14} that the average lifetime $\tau$
grows as $\exp(\mathcal{S}^\omega)$, with $\omega\approx \frac{1}{2}$ in
$d=3$ and with $\omega$ slowly growing with $d$.  Simulation results up to 5
dimensions suggest that $\beta$ is a continuously growing function of
dimension~\cite{BR14} and $\beta$ reaches the mean-field value of 1 only when
$d\to\infty$.  Preliminary numerical simulations indicate that the
distribution of times between visits to distinct sites has a slower than
exponential decay for spatial dimensions $d>2$ and only gradually approaches
an exponential decay for $d\to\infty$~\cite{todo}.  These features suggest
that the critical dimension of the starving random walk problem may be
infinite.

Thus starving random walks represent a new type of non-equilibrium process in
which the dynamics is controlled by the times \emph{between visits to
  distinct sites} in a random walk.  While much is known about the average
number of distinct sites visited and its underlying distribution for random
walks~\cite{F68,W94}, little appears to be known about the times between
visits to new sites, as well as the distribution of times between such
visits.  This aspect of the starving random walk model is very much worth
exploring.
\section*{Acknowledgments}

MC and OB acknowledge support from the European Research Council starting
grant No.\ FPTOpt-277998.  Financial support for this research was also
provided in part by the grant DMR-1623243 from the National Science
Foundation and by a grant from the John Templeton Foundation.

\appendix

\section{Distribution of the number of distinct sites
  visited in one dimension}
\label{P}

For a random walk that starts at $x=a$ within an interval of length $L=ka$
with both ends absorbing, the concentration at position $x$ at time $t$ is
\begin{equation}
c(x,t) = \frac{2}{ka}\,\sum_{n\geq 1} \sin\frac{n\pi x}{ka}\,\sin\frac{n\pi }{k}\, 
\exp\left[-\left(\frac{n\pi}{ka}\right)^2 Dt\right]\,.
\end{equation}
The flux leaving the interval at time $t$ is therefore
\begin{align}
  F_k(t)&= D\left(c'\big|_{x=0} - c'\big|_{x=ka}\right)\,\nonumber\\
  &=\frac{2D}{ka}\,\sum_{n=1}^\infty \frac{n\pi}{ka} \,\sin\frac{n\pi}{k}\,
(1-\cos n\pi)\,\exp\left[-\left(\frac{n\pi}{ka}\right)^2 Dt\right]\,,\nonumber \\
  &=\frac{4\pi D}{(ka)^2}\sum_{n=0}^\infty (2n+1)\, \sin\frac{(2n+1)\pi}{k}\,
  \exp\left[-\left(\frac{(2n+1)\pi}{ka}\right)^2 Dt\right]\,.
\end{align}
We rewrite the sum over odd integers $n$ as the sum over all integers $2n+1$.
Thus the integrated flux that leaves the interval up to time $\mathcal{S}$ is
\begin{align}
\mathcal{F}_k    = 1-\frac{4}{\pi}\sum_{n\geq 0} \frac{1}{2n+1} \,\sin\frac{(2n+1)\pi}{k}\, 
\exp\left[-\left(\frac{(2n+1)\pi}{ka}\right)^2 D\mathcal{S}\right]\,.
\end{align}

We now need to compute the product
$\prod_{2\leq k\leq \mathcal{N}} \mathcal{F}_k\equiv U(\mathcal{N})$ that
appears in \eqref{PLdef}.  Taking the logarithm, we have
\begin{equation}
  \ln U(\mathcal{N}) = \sum_{k=2}^\mathcal{N} \ln\left\{ 1 -
    \frac{4}{\pi}\sum_{n}\frac{\sin[(2n+1)\pi/k]}{2n+1}\, 
    \exp\left[- \big((2n+1)\pi\big)^2 D\mathcal{S}/(ka)^2\right]\right\}\,.
\end{equation}
We convert the sum to an integral, introduce the scaled variable
$z=n/\sqrt{\mathcal{S}}$, and replace $\sin[(2n+1)\pi/k]/(2n+1)$ by $\pi/k$;
this latter approximation applies in the limit of large $\mathcal{S}$.  These
steps lead to
\begin{equation}
\ln U(\mathcal{N}) \simeq \sqrt{\mathcal{S}}\int_0^{\mathcal{N}/\sqrt{\mathcal{S}}} dz\, 
\ln\left\{1-\frac{2}{\pi}\,\frac{2\pi}{z\sqrt{\mathcal{S}}}\sum_{n\geq 0} 
\exp\left[-\frac{\big((2n+1)\pi\big)^2 D}{(az)^2}\right]\right\}\,.
\end{equation}
We expand the logarithm, which again applies for large $\mathcal{S}$, and
ultimately obtain
\begin{align}
\ln U(\mathcal{N}) &\simeq - 4 \int_0^{\mathcal{N}/\sqrt{\mathcal{S}}}
\frac{dz}{z}\,\sum_{n\geq 0} \exp\left[-\frac{\big((2n+1)\pi\big)^2 D}{(az)^2}\right]\,,\nonumber \\
&= - 2\sum_{n\geq 0} \mathrm{E}_1\big((2n+1)^2/\theta^2\big)\,,
\end{align}
where $\theta \equiv \mathcal{N}a/(\pi\sqrt{D\mathcal{S}})$ and
E$_1(x)\equiv \int_1^{\infty} dt \, e^{-xt}/t$ denotes the exponential
integral.

Similarly, the factor $1-\mathcal{F}_{\mathcal{N}+1}$ in \eqref{PLdef} has
the $\mathcal{S}\to\infty$ asymptotic behavior
\begin{equation}
1-\mathcal{F}_{\mathcal{N}+1}\simeq \frac{2}{\pi}\sum_{n\geq 0}
\frac{2a}{\theta\sqrt{D\mathcal{S}}}\, e^{-(2n+1)^2/\theta^2}\,.
\end{equation}
Finally, using Eq.~\eqref{PLdef}, the distribution of the scaled variable $\theta$ is
\begin{equation}
\label{Q}
V(\theta) = \frac{4}{\theta} \exp\left[ -2\sum_{n\geq 0} \mathrm{E}_1\big((2n+1)^2/\theta^2\big)\right] \sum_{n\geq 0} e^{-(2n+1)^2/\theta^2}
\,.
\end{equation}
We use this form in Eq.~\eqref{N} to compute the average number of distinct
sites visited at the instant of starvation.

\section{Average lifetime $\tau$ in one dimension}
\label{tau}

Using the results of the \ref{P}, we write the numerator in
Eq.~\eqref{tau-n} as
\begin{equation}
  \mathcal{G}_k\equiv \int_0^\mathcal{S} dt\,\, t \,\, F_k(t) = \frac{4\pi
    D}{(ka)^2}\sum_{j=0}^\infty   (2j+1)\sin\left[\frac{(2j+1)\pi}{k}\right]
  \int_0^\mathcal{S} dt\,\, t\,\, e^{-\beta t}\,,
\end{equation}
with $\beta\equiv \big[(2j+1)\pi\big]^2 D/(ka)^2$.  For large $k$, we
approximate the sine function by its argument and then perform the temporal
integral to give
\begin{align}
\label{Gk}
  \mathcal{G}_k&\simeq \frac{4\pi D}{(ka)^2}\, \frac{\pi}{k}\sum_{j=0}^\infty (2j+1)^2
                 \left[\frac{k^2 a^2}{(2j+1)^2\pi^2 D}\right]^2 \left[1-
                 e^{-\beta\mathcal{S}}(1+\beta\mathcal{S})\right]\,,\nonumber \\
               &  \simeq \frac{4ka^2}{\pi^2 D}\sum_{j=0}^\infty \frac{1}{(2j+1)^2}\left[1-
                 e^{-\beta\mathcal{S}}(1+\beta\mathcal{S})\right]\,.
\end{align}
To leading order in $\mathcal{S}$, the denominator in \eqref{tau-n} is
subdominant, so that we have $\tau_k \simeq \mathcal{G}_k.$ We now define
$\mathcal{T}_n\equiv \sum\limits_{k=1}^n \tau_k$.  Using Eq.~\eqref{Gk}, we
again convert the sum over $k$ to an integral and introduce the scaled
variable $u=ka/(\pi \sqrt{D\mathcal{S}})$ to give
\begin{align}
\mathcal{T}_n \simeq \mathcal{S} \int_0^{na/(\pi \sqrt{D\mathcal{S}})} du \, u 
\sum_{j=0}^\infty \frac{4}{(2j+1)^2}\left\{1-e^{-(2j+1)^2/u^2}\left[1+\left(\frac{2j+1}{u}\right)^2\right]\right\}\,.
\end{align}
From Eq.~\eqref{tausum}, we write
\begin{equation}
\tau = \sum\limits_{n=0}^{\infty} \mathcal{T}_n V(n)+\mathcal{S}\,.
\end{equation}
Introducing $\theta=na/(\pi \sqrt{D\mathcal{S}})$ and replacing the sum over
$n$ by an integral, we obtain
 \begin{align}
 \tau &\simeq \mathcal{S} \int_0^{\infty}\! \!d\theta \, V(\theta)\int_0^{\theta}\! du \, u 
\sum_{j=0}^\infty \frac{4}{(2j+1)^2}\left\{\!1\!-\!e^{-(2j+1)^2/u^2}\left[1+\left(\frac{2j\!+\!1}{u}\right)^2\right]\right\}\! +\!\mathcal{S}\,,
 \end{align}
 and using Eq.~\eqref{Q}, the average lifetime is $\tau \simeq 3.27686\ldots \, \mathcal{S}$.

\section{The distribution $V(\mathcal{N})$ in the circular approximation}
\label{P2D}

Consider a random walk that starts a distance $a$ from the boundary of a
circle of radius $R_k\equiv\sqrt{k/\pi}\,a$.  For $k\leq 3$, the radius is
smaller than $a$, so for these cases the walk starts from the center of the
circle.  We neglect this detail in the following.  The probability to first
reach the boundary at time $t$ is given by~\cite{R01}
\begin{equation}
F_k(t)=\frac{2D}{R_k^2} \sum\limits_{m=1}^{\infty} j_m \,e^{-D t j_m^2/R_k^2}\,\, 
\mathrm{J}_0\bigg(j_m \Big( 1-\frac{a}{R_k} \Big) \bigg)\Big/ \mathrm{J}_1(j_m)\,,
\end{equation}
with J$_0$ and J$_1$ the Bessel functions of order 0 and 1, and $j_m$ the
$m^{\rm th}$ zero of J$_0$.  Integrating this expression for $0\leq t\leq
\mathcal{S}$, the probability $\mathcal{F}_k$ to reach the boundary before
starving is then\footnote{Eq.~\eqref{corr} corrects a misprint in the corresponding equation in~\cite{BR14}.}
\begin{equation}\label{corr}
\mathcal{F}_k= 1-2\sum\limits_{m=1}^{+\infty} e^{-D\mathcal{S}j_m^2/R_k^2}\,\,
\mathrm{J}_0\Big(j_m \big( 1-\sqrt{{\pi}/{k}} \big) \Big)\Big/\big[j_m \, \mathrm{J}_1(j_m)\big]\, .
\end{equation}
As in one dimension, we need
$U(\mathcal{N})=\prod_{1\leq k\leq \mathcal{N}} \mathcal{F}_k$.  We start by
taking its logarithm
\begin{equation}
\ln U(\mathcal{N}) = \sum\limits_{k=1}^{\mathcal{N}} \ln \left\{
  1-2\sum\limits_{m=1}^{+\infty} 
e^{-D\mathcal{S}j_m^2/R_k^2}\,\, \mathrm{J}_0\left(j_m
  \big(1-\sqrt{{\pi}/{k}}\big) \right)\Big/ \big[j_m \, \mathrm{J}_1(j_m)\big] \right\}\,.
\end{equation}
We again convert the sum to an integral, expand the logarithm and use
\begin{equation}
\label{J-identity}
\frac{\mathrm{J}_0\left(j_m \big( 1-\sqrt{{\pi}/{k}} \big) \right)}{\mathrm{J}_1(j_m)} \sim j_m \sqrt{\frac{\pi}{k}}
\end{equation}
when $\mathcal{S}$ and consequently $k$ become large.  This yields
\begin{equation}
\ln U(\mathcal{N}) \simeq -4 \sqrt{\pi \mathcal{N}} \sum\limits_{m=1}^{+\infty} 
\left[ e^{-D\pi \mathcal{S}j_m^2/(\mathcal{N} a^2)}  
-\sqrt{\frac{D \mathcal{S}}{\mathcal{N}}} \, \frac{\pi j_m}{a} \, 
{\rm erfc}\Big({\sqrt{\frac{D\pi \mathcal{S}}{\mathcal{N}}} \frac{j_m}{a}} \Big) \right] \,.
\end{equation}
Similarly, the term $1-\mathcal{F}_{\mathcal{N}+1}$ has the following
asymptotic behavior in the large $\mathcal{S}$ limit
\begin{equation}
1-\mathcal{F}_{\mathcal{N}+1} \simeq  \sqrt{4\frac{ \pi }{\mathcal{N}}} 
\sum\limits_{k=1}^{+\infty} e^{-D\pi \mathcal{S}j_k^2/(\mathcal{N} a^2) }\,.
\end{equation}
Finally, the distribution of the number of distinct sites visited at
starvation is given by
\begin{align} 
V(\mathcal{N}) \simeq  &\sqrt{\frac{4 \pi }{\mathcal{N}}} 
\sum\limits_{k=1}^{+\infty} e^{-D\pi \mathcal{S}j_k^2/(\mathcal{N} a^2)}  \nonumber \\
& \times \exp \left\{\!-4 \sqrt{\pi \mathcal{N}} \sum\limits_{m=1}^{+\infty} 
\!\left[ e^{-D\pi \mathcal{S}j_m^2/(\mathcal{N} a^2)} 
-\sqrt{\frac{D \mathcal{S}}{\mathcal{N}}} \, \frac{\pi j_m}{a} \, 
{\rm erfc} \Big( {\sqrt{\frac{D\pi \mathcal{S}}{\mathcal{N}}} \frac{j_m}{a}} \Big)\! \right]\! \right\} \,.
\end{align}
Introducing the variable $\theta= \mathcal{N} a^2/(\pi D \mathcal{S})$, we obtain
Eq.~\eqref{Pcirc}.

\section{Average lifetime in the circular approximation}
\label{tau2D}

As in one dimension, we write the numerator in Eq.~\eqref{tau-n} as
\begin{equation}
  \mathcal{G}_k\equiv \int_0^\mathcal{S}\!\! dt\,\, t \,\, F_k(t) =
  \frac{2D\pi}{ka^2} \int_0^{\mathcal{S}} \!\!dt \, t \sum\limits_{m=1}^{\infty}  j_m 
\frac{\mathrm{J}_0\left(j_m \big( 1-\sqrt{\pi/k} \big) \right)}{\mathrm{J}_1(j_m)} 
\,\, e^{-\pi D j_m^2 t/(k a^2)} \,,
\end{equation}
with $j_m$ the $m^{\rm th}$ zero of the Bessel function J$_0$.  For large
$k$, we again substitute \eqref{J-identity} in the above integrand
and then perform the temporal integral to give
\begin{equation}
\label{Gn}
  \mathcal{G}_k\simeq \frac{2 a^2}{D} \, \sqrt{\frac{k}{\pi}}
 \sum\limits_{m=1}^{\infty} \frac{1}{j_m^2} 
\left[ 1 - \left( 1+\frac{\pi D j_m^2 \mathcal{S}}{k a^2} \right) 
e^{-\pi D \mathcal{S}j_m^2/(k a^2)} \right]\,.
\end{equation}
To leading order in $\mathcal{S}$, we still have
$\tau_k \simeq \mathcal{G}_k$.  We then define
$\mathcal{T}_n \equiv \sum\limits_{k=1}^n \tau_k$, use
$\tau_k \simeq \mathcal{G}_k$, with $\mathcal{G}_k$ given in \eqref{Gn}, and
convert the sum to an integral to give
\begin{equation}
\mathcal{T}_n \simeq \frac{2 a^2}{D} \int_0^n dk \sqrt{\frac{k}{\pi}} 
\sum\limits_{m=1}^{\infty} \frac{1}{j_m^2} 
\left[ 1 - \left( 1+\frac{\pi D j_m^2 \mathcal{S}}{k a^2} \right) 
e^{-\pi D \mathcal{S}j_m^2/(k a^2)} \right] \,.
\end{equation}
Changing to the variable $u=k a^2/(\pi D \mathcal{S})$ leads to
\begin{equation}
\mathcal{T}(\theta) \simeq \frac{2 \pi \sqrt{D}\, \mathcal{S}^{3/2}}{a}
\int_0^{\theta} \frac{du}{\sqrt{u}} \sum\limits_{m=1}^{\infty}  
\left[ \frac{u}{j_m^2} - \left( 1+\frac{u}{j_m^2} \right)
e^{-j_m^2/u} \right] \,,
\end{equation}
where $\theta\equiv n a^2/(\pi D \mathcal{S})$.  Using Eq.~\eqref{tausum}, we write
\begin{equation}
\tau = \sum\limits_{n=0}^{\infty} \mathcal{T}_n V(n)+\mathcal{S} 
\simeq \int_0^{\infty} \mathcal{T}(\theta) V(\theta,\mathcal{S}) +\mathcal{S}\,.
\end{equation}
with $V(\theta,\mathcal{S})$ given in Eq.~\eqref{Pcirc}. We finally obtain
 \begin{align}
 \tau &\simeq \mathcal{S} +\frac{2 \pi \sqrt{D} \,\mathcal{S}^{3/2}}{a}  
\int_0^{\infty} d\theta \,\, V(\theta,\mathcal{S}) 
\int_0^{\theta} \frac{du}{\sqrt{u}} \sum\limits_{m=1}^{\infty}  
\left[ \frac{u}{j_m^2} - \left( 1+\frac{u}{j_m^2} \right) e^{-{j_m^2}/{u}} \right] \,.
 \end{align}

\end{document}